# Bimodal coalitions and neural networks

L.B. Litinskii, and I.M. Kaganowa

Center of Optical Neural Technologies,
Scientific Research Institute for System Analysis of Russian Academy of Sciences,
Nakhimov ave, 36-1, Moscow, 117218, Russia
*litin@mail.ru*, *imkaganova@gmail.com*

## Abstract

We give an account of the Axelrod – Bennet model that describes formation of a bimodal coalition. We present its initial formalism and applications and reformulate the problem in terms of the Hopfield model. This allowed us to analyze a system of two homogeneous groups of agents, which interact with each other. We obtained a phase diagram describing the dependence of the bimodal coalition on external parameters.

## Introduction

In the early 90s R. Axelrod and D. Bennet proposed an approach for a formal description of splitting of a set of interacting agents into two competing groups [1], [2]. Their results have found applications in social, politic, and management sciences. Then Serge Galam [3] reformulated this approach in terms of the Ising model. Afterwards he complicated the initial scheme and proposed a number of new models (see references in [4]). The following development of this approach led to the appearance of the econophysics and sociophysics.

In this paper, we solve the same problem using the ideas and concepts of the discrete dynamic of the Hopfield model. We analyze analytically an idealized case of two equally interacting homogeneous groups of the agents and construct a phase diagram that describes completely how the decomposition of the agents into two groups depends on the intra-group interaction and cross-interaction between groups.

Following tradition, the decomposition of the agents into two groups will be called *a bimodal coalition*.

## Bimodal coalition problem

**1. The original setting of the problem.** We have $n$ agents that are connected with each other. By $w_i$, $i=1,...,n$ we define the weight of the $i$-th agent. The connections of the agents we interpret in terms of their mutual propensity and suppose that propensities are symmetrical:

$$p_{ij} : \begin{cases} > 0, & \text{if agents } i \text{ and } j \text{ are prone to cooperate;} \\ < 0, & \text{if agents } i \text{ and } j \text{ are prone to conflict.} \end{cases} \qquad p_{ij} = p_{ji}.$$

Two lists $A$ and $\tilde{A}$ define a bimodal coalition $C = (A, \tilde{A})$ or, in other words, *decomposition into two groups*. Each of these lists contains all the numbers of agents assigned to the given group:

$$A = \{i_1, i_2, ..., i_p\}, \; \tilde{A} = I \setminus A, \text{ where } I = \{1, 2, ..., n\} \text{ is the full list.}$$

Each grouping $C = (A, \tilde{A})$ provides a proximity relation $d_{ij}$ between the agents:

$$d_{ij}(C) = \begin{cases} 1, & \text{if agents } i \text{ and } j \text{ belong to the given list;} \\ 0, & \text{if agents } i \text{ and } j \text{ belong to different lists.} \end{cases}$$

Let us define a productivity of the grouping $C = (A, \tilde{A})$ for the $i$-agent as

$$U_i(C) = \sum_{j=1}^{n} w_j p_{ij} d_{ij}(C).$$

The productivity of the grouping $C$ for the $i$-th agent is maximal if all the other agents with which the given agent is prone to cooperate belong to his group and the group does not contain agents with which it is prone to conflict.

In the Axelrod-Bennet it is stated that a system of agents tends to those grouping for which the weighted sum of the productivities is maximal:

$$U(C) = \sum_{i=1}^{n} w_i \cdot U_i(C) \to \max. \tag{1}$$

**2. Applications.** The described approach was applied when analyzing compositions of the belligerent coalitions during the World War II. The agents were 17 European countries. An integral index defined the weight of each country. It was calculated as a combination of different demographic, industrial, and military characteristics.

The mutual propensities were calculated using the data for 1936 and criteria that included ethnic conflicts, religion, frontier incidents, political regime and so on. The maximization of the sum $U(C)$ led to two maximums. The global maximum corresponded to the following decomposition:

$A$ = {Britain, France, USSR, Czechoslovakia, Yugoslavia, Greece, and Denmark};
$\tilde{A}$ = {Germany, Italy, Poland, Romania, Hungary, Portugal, Finland, Latvia, Lithuania, and Estonia}.

We see that only Poland found itself in the improper camp (as well as Portugal, which was a neutral nation during the war). Let us make it clear that the block to which the given country belonged was determined by taking into account who occupied the country or who declared war on it.

In other paper of the same authors they used this method to describe alliances of produsers of standards of UNIX operating systems. Nine companies involved in the UNIX production were regarded as agents. They are

AT&T, Sun, Apollo, DEC, HP, Intergraph, SGI, IBM and Prime.

In the course of cumbersome calculations of the connections $p_{ij}$, some parameters of the problem played the role of weight coefficients. By varying the parameters within reasonable limits they discovered only a weak dependence of the result on the values of the parameters. The authors found that there were two decompositions of the functional (1) that provided the same global maximum:

- {Sun, DEC, HP} and {AT&T, Apollo, Intergraph, SGI, IBM, Prime};
- {Sun, AT&T, IBM, Prime} and {DEC, HP, Apollo, Intergraph, SGI}.

The second grouping corresponded to the existing associations of the companies in UNIX International and OPEN Software Foundation and only IBM was identified incorrectly.

**3. Ising model.** In the second half of 90s Serge Galam recognized that it was convenient to formulate the Axelrod-Bennet model in terms of the Ising model.

Let us introduce a matrix

$$\mathbf{J} = (J_{ij}), \ J_{ij} = p_{ij} w_i w_j (1 - \delta_{ij}), \ \ i, j = 1, n$$

where $\delta_{ij}$ is the Kronecker delta and the diagonal elements of the matrix $\mathbf{J}$ are equal to zero. To each bimodal coalition $C$ we assign a configuration vector $\mathbf{s} = (s_1, s_2, ..., s_n)$:

$$C = (A, \tilde{A}) \ \Leftrightarrow \ \mathbf{s} = (s_1, s_2, ..., s_n): \ \begin{cases} s_i = 1, & i \in A, \\ s_i = -1, & i \in \tilde{A}. \end{cases}$$

Then the maximization of the sum (1) is equivalent to the determination of the state $\mathbf{s}$ corresponding to the global minimum of the energy $E(\mathbf{s})$:

$$E(\mathbf{s}) = -(\mathbf{Js}, \mathbf{s}) = -\sum_{i,j=1}^{n} J_{ij} s_i s_j \to \min. \tag{2}$$

The problem (2) is a well-known minimization problem of a quadratic form of binary variables. This problem arises in various scientific fields.

**4. Hopfield model.** Let us analyze the described system in terms of a neural network of the Hopfield type. As the context may require, in what follows we refer to the binary variables $s_i = \pm 1$ as *binary agents* or *spins*. The state of the system we describe by a configuration vector $\mathbf{s} = (s_1, s_2, ..., s_n)$.

Let us introduce a dynamic procedure on which the Hopfield model is based. *Let $\mathbf{s}(t)$ be the state of the system at time $t$. At this moment a local field $h_i(t) = \sum_{j=1}^{N} J_{ij} s_j(t)$ acts on the i-th spin. At the next moment $t+1$ the state of the spin changes if its sign does not coincide with the sign of the field $h_i(t)$, and it remains unchanged otherwise*:

$$s_i(t+1) = \begin{cases} s_i(t), & \text{when } s_i(t)h_i(t) \geq 0 \\ -s_i(t), & \text{when } s_i(t)h_i(t) < 0 \end{cases} \Leftrightarrow s_i(t+1) = sign(h_i(t)). \qquad (3)$$

In what follows, an *unsatisfied* spin is a spin whose sign does not coincide with the sign of the field acting on it. If the state of the *i*-th spin changes, then its contribution to the local fields acting on the other spins also changes. As a result, the state of some other spins can also change etc. The evolution of the system consists of subsequent turns of unsatisfied spins. Each step of the evolution is accompanied by a decrease of the energy of the state, and sooner or later the system reaches a state that corresponds to an energy minimum (it may be a local minimum). At that moment, the evolution of the system will stop, since all the spins will be satisfied. However, according the setting of the problem, we have to find the global minimum. For this purpose we can use improved procedures of minimization [5], [6]. The formulation of problem (3) in terms of neural networks allows us to illustrate the problem of bimodal coalition formation.

Concluding this section let us note, that all the energies are two-fold degenerate: $E(\mathbf{s}) = -(\mathbf{Js},\mathbf{s}) = E(-\mathbf{s})$. To remove the degeneration we need an external field.

### Homogeneous groups of agents

**1. One homogeneous group.** A homogeneous group is a group where all the agents interact identically. In this case the interaction matrix has the form

$$\mathbf{J} = \begin{pmatrix} 0 & a & \cdots & a \\ a & 0 & \cdots & a \\ \vdots & \vdots & \ddots & \vdots \\ a & a & \cdots & 0 \end{pmatrix}, \quad a > 0. \qquad (4)$$

The network with such a connection matrix has only one the global minimum of the energy $\mathbf{s}_0 = (1,1,...,1)$, and there are no other minima. (We do not take into account the second minimum that appears due to the equality $E(\mathbf{s}) = E(-\mathbf{s})$). In other words, the states of all the agents are the same. It can be said that all the agents behave "as one person".

If we turn to equation (2) we see that for the system with the connection matrix (4), not a bimodal coalition but a consolidation of all the agents into one group is profitable.

**2. Two homogeneous groups.** Let us examine a spin system consisting of two homogeneous groups. We suppose that in the first group there are *p* agents and the interactions between these agents are identical and equal to *A*. The interactions between the remaining *q* agents (that constitute the second group) are also identical and equal to *C*. We suppose that all the interactions between the agents from the first and second groups are equal to B. We assume that C is positive and larger than A and B, and factor out C. Now the connection matrix has the form

$$\mathbf{J} = \begin{pmatrix} \overbrace{0 \quad a \quad \cdots \quad a}^{p} & \overbrace{b \quad b \quad \cdots \quad b}^{q} \\ a \quad 0 \quad \cdots \quad a & b \quad b \quad \cdots \quad b \\ \vdots \quad \vdots \quad \ddots \quad \vdots & \vdots \quad \vdots \quad \ddots \quad \vdots \\ a \quad a \quad \cdots \quad 0 & b \quad b \quad \cdots \quad b \\ b \quad b \quad \cdots \quad b & 0 \quad 1 \quad \cdots \quad 1 \\ b \quad b \quad \cdots \quad b & 1 \quad 0 \quad \cdots \quad 1 \\ \vdots \quad \vdots \quad \ddots \quad \vdots & \vdots \quad \vdots \quad \ddots \quad \vdots \\ b \quad b \quad \cdots \quad b & 1 \quad 1 \quad \cdots \quad 0 \end{pmatrix}, \quad \begin{matrix} a = A/C, \\ b = B/C. \end{matrix}$$

For a neural network with such a connection matrix we will describe the dependence of the set of the minima upon the parameters $a$, $b$, $p$, and $q$, where $p+q=n$.

One can show that if a configuration corresponds to a minimum of the energy (2) its last $q$ coordinates have to be identical:

$$\mathbf{s} \sim (s_1, s_2, ..., s_p, 1, 1, ..., 1). \tag{5}$$

Consequently, there are at most $2^p$ configurations that can be minima of the functional (2) and the last $q$ coordinates of all the configurations are identical.

Let us divide the set of $2^p$ configurations into classes $\Sigma_k$ in such a way that the configurations where among the first $p$ coordinates exactly k coordinates equal to $-1$ belong to the class $\Sigma_k$. Since $k$ takes the values $k = 0, 1, 2, ..., p$, there are $p+1$ such classes. Let us write down these classes in the explicit form indicating how many configurations belong to a given class (see Table 1).

**Table 1.** Classes $\Sigma_k$ and numbers of configurations in these classes.

| | |
|---|---|
| $\Sigma_0 = (\underbrace{1, 1 ... 1}_{p}, \underbrace{1, ... 1}_{q})$ | unique configuration |
| $\Sigma_1 = \begin{cases} (\mathbf{-1}, 1, ..., 1, 1, ... 1) \\ (1, \mathbf{-1}, ..., 1, 1, ... 1) \\ ... \\ (\underbrace{1, 1, ..., \mathbf{-1}}_{p}, \underbrace{1, ... 1}_{q}) \end{cases}$ | $p$ configurations |
| ……. | …… |
| $\Sigma_k = \left\{ \mathbf{s} : \sum_{i=1}^{p} s_i = p - 2k \right\}$ | $C_p^k$ configurations |
| …….. | ……. |
| $\Sigma_p = (\underbrace{-1, -1 ..., -1}_{p}, \underbrace{1, ... 1}_{q})$ | unique configuration |

It turns out that for some values of the parameters, only the configurations from the class $\Sigma_k$ provide the minimum of the functional (2) simultaneously. For all the configurations from $\Sigma_k$ the energies (2) are the same. In other words, they are minima, and if the inequality $0 < k < p$ is fulfilled then there are no other local minima of the functional (2).

In Fig.1 we show the partition of the $(a,b)$-plane into regions where one class or other of the configurations $\Sigma_k$ provides a minimum of the functional (2). Below we interpret this diagram in terms of bimodal coalitions.

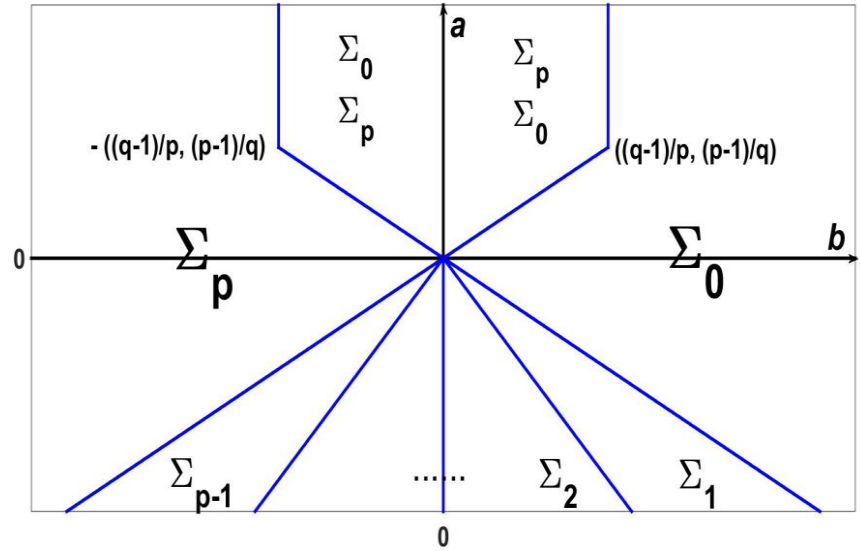

**Fig. 1.** The phase diagram for the problem (2).

**3. Sensible interpretation.** From Eq. (5) for the coordinates of the global minimum it follows that the second group always acts "as one person" – the last $q$ spins are equal to +1.

At first, let us examine the case when the agents of the first homogeneous group are prone to cooperate with each other ($a > 0$). Then they also act 'as one person' (see the upper half-plane of the diagram). If both groups of agents are prone to cooperate with each other, that is, when $b > 0$, then all the agents of the first group are in the same state as the agents of the second group, i.e. the first $p$ coordinates of the vector $\Sigma_0$ are equal to +1. However, if the groups conflict with each other, that is, $b < 0$, then all the agents of the first group are in the state, opposite to the state of the agents belonging to the second group.

Let us summarize. When all the agents inside each group are prone to cooperate ($a > 0$), the sign of the cross-interaction defines the state of the whole system. If $b > 0$, and, consequently, the groups are prone to cooperate with each other, it is more profitable for them to be together. In this case, the vector $\Sigma_0$ provides the global minimum. If $b < 0$ and the groups are conflicting, it is more profitable for the groups to be separate: in this case the global minimum corresponds to $\Sigma_p$.

Inside the symmetric strip along the axis of ordinates, both configurations $\Sigma_0$ and $\Sigma_p$ are minima simultaneously. This strip is a unique region on the plane where the functional (2) has both global and local minima simultaneously. To the right of the axis of ordinates, where $b > 0$, the vectors $\Sigma_0$ and $\Sigma_p$ provide the global and the local minima, respectively. On the other hand, to the left of the axis of ordinates, $\Sigma_p$ corresponds to the global minimum and $\Sigma_0$ to the local minimum. It is easy to explain why such quasi-instability takes place. Indeed, let us suppose that the cross-interaction between the groups is equal to zero: $b = 0$. In other words, two groups of the agents are completely independent. Then the problem (2) has two equivalent solutions $\Sigma_0$ and $\Sigma_p$ that correspond to the same value of energy. When $|b|$ increases slightly, at the beginning the second configuration continues to be a minimum, but now a local minimum. When the value of $|b|$ becomes sufficiently large, the additional local minimum disappears.

The narrow strip along the axis of ordinates is the result of removing the random degeneracy of the global minimum when the external parameter $b = 0$. It is interesting to understand whether local minima always appear for the same reason? Or are there other mechanisms for their appearance?

Finally, let us briefly discuss the situation when the agents inside the first group conflict with each other ($a < 0$). The lower half of the phase diagram shows that in this case the first group of agents splits into two opposing groups. This conclusion is rather reasonable. Other intrinsic interpretations are more speculative.

## Conclusions

We have shown that in a system with a great number of interacting binary agents, the known problem of the formation of two competing groups, or the problem of the bimodal coalition, can be formulated in terms of neural networks of the Hopfield type. The neural network dynamics is convenient when describing the influence of the agents on each other. We analyzed theoretically an idealized case of interaction between two homogeneous groups of agents. The obtained results allowed us to present a sensible interpretation of the bimodal coalition problem.

We determined the mechanism of the formation of the local minima for the energy functional. It is interesting to find out whether there are other possibilities for their appearance. We think that our analysis is promising and deserves further examination.

The work was financially supported by State Program of SRISA RAS No. 0065-2019-0003 (AAA-A19-119011590090-2). We are grateful to Ben Rozonoer for his help in preparation of this paper.

## References

1. Axelrod R M and Bennett D S. A Landscape Theory of Aggregation // British Journal of Political Science. 1993. Vol. 23/ No. 2. P. 211-233.

2. Axelrod R M, Mitchell W, Thomas R E, Bennett D S, and Bruderer E. Coalition Formation in Standard-Setting Alliances // Management Science. 1995. Vol. 41. No. 9. P. 1493-1508.

3. Galam S. Fragmentation versus stability in bimodal coalitions // Physica A. 1996. Vol. 230. No. 1-2. P. 174-188.

4. Galam Serge. Sociophysics. New York: Springer-Verlag. 2012. 270p.
5. Houdayer J., Martin O. C. Renormalization for Discrete Optimization // Phys. Rev. Lett. 1999. Vol. 83. P. 1030-1033.
6. Karandashev I. M. and Kryzhanovsky B. V. Matrix Transformation Method in Quadratic Binary Optimization // Optical Memory and Neural Networks (Information Optics). 2015. Vol. 24. No. 2. P. 67-81.